\documentclass[12pt]{article}
\usepackage{amssymb,amsmath,amscd,amsthm}

\newtheorem{theorem}{Theorem}[section]

\newtheorem{lemma}[theorem]{Lemma}

\newcommand{\sol}{u^{\text{sc}}}
\newcommand{\obs}{\mathcal S}

\date{}

\begin{document}
 \author{ Evgeny Lakshtanov\thanks{Department of Mathematics, Aveiro University, Aveiro 3810, Portugal.  This work was supported by Portuguese funds through the CIDMA - Center for Research and Development in Mathematics and Applications and the Portuguese Foundation for Science and Technology (``FCT--Fund\c{c}\~{a}o para a Ci\^{e}ncia e a Tecnologia''), within project PEst-OE/MAT/UI4106/2014 (lakshtanov@ua.pt).} \and
 Boris Vainberg\thanks{Department
of Mathematics and Statistics, University of North Carolina,
Charlotte, NC 28223, USA. The work was partially supported  by the NSF grant DMS-1410547 (brvainbe@uncc.edu).}}

\title{Recovery of interior eigenvalues from reduced near field data }

 \maketitle
\begin{abstract}
We consider inverse obstacle and transmission scattering problems where the source of the incident waves is located on a smooth closed surface that is a boundary of a domain located outside of the obstacle/inhomogeneity of the media. The domain can be arbitrarily small but fixed.The scattered waves are measured on the same surface. An effective procedure is suggested for recovery of interior eigenvalues by these data.
\end{abstract}



\section{Introduction}
Consider bounded, connected domain $\mathcal O \subset \mathbb R^d, d=2,3$,  with twice differentiable boundary $\partial \mathcal O$. The scattering
problem is stated as either an obstacle scattering problem
\begin{equation}\label{eq1}
\begin{array}{l}
-\Delta \sol -\lambda \sol =0, \quad x \in \mathbb R^d\backslash\mathcal O, ~ \lambda=k^2>0,\\
\sol+u^{inc}=0, \quad x \in \partial \mathcal O, \\
\end{array}
\end{equation}
or a transmission scattering problem
\begin{equation}\label{eq2}
-\Delta \sol -\lambda n(x)\sol =\lambda [1-n(x)]u^{inc}, \quad x \in \mathbb R^d, ~ \lambda=k^2>0, \\
\end{equation}
where $n(x)=1, ~x\in \mathbb R^d\backslash\overline{\mathcal O},~n(x)>0, ~x\in \overline{\mathcal O}, ~n$ is smooth in $\mathcal O$ up to the boundary $\partial\mathcal O$, and has interior limiting values at the boundary that are not equal to one, i.e., $n(x)-1$ has a constant sign when $x \in \partial \mathcal O$. Here $u^{inc}$ is an incident wave that satisfies the Helmholtz equation in $\mathbb R^d$ or $\mathbb R^d \backslash \obs$.  Here $\obs$ is a bounded set where the sources are distributed. We will specify $\obs$ later. Solution $\sol$ satisfies the radiation condition:
\begin{equation}\label{rc}
\sol=\sol_\infty(k,\theta)\frac{e^{ikr}}{r^{\frac{d-1}{2}}}  + O \left (r^{-\frac{d+1}{2}} \right ),\quad \theta=\frac{x}{r},~~r=|x|\to\infty.
\end{equation}
The Dirichlet boundary condition in (\ref{eq1})  can be replaced by the Neumann boundary condition. The transmission problem concerns scattering by inhomogeneous media. The media may include an obstacle $\mathcal O_1$ inside of the region of inhomogeneity.  Then equation (\ref{eq2}) holds in $\mathbb R^d\backslash\overline{\mathcal O_1}$ with the corresponding boundary condition on $\partial \mathcal O_1$, see e.g. \cite{cch}. We will assume that $\mathcal O_1=\emptyset$, but all the results below can be automatically extended to the case $\mathcal O_1\neq\emptyset$.

{\it Interior eigenvalues} are the Dirichlet/Neumann negative Laplacian eigenvalues in $\mathcal O$ for the first problem and interior transmission eigenvalues (ITEs) for the second problem. The latter are defined as the values of $\lambda$ for which the following system of equations in $\mathcal O$
\begin{equation}\label{Anone0}
-\Delta u - \lambda u =0, \quad -\Delta v - \lambda   n(x)v =0, \quad x \in \mathcal O, \quad u,v\in H^2(\mathcal O),
\end{equation}
\begin{equation}\label{Antwo}
\begin{array}{l}
u-v=0, \quad x \in \partial \mathcal O, \\
\frac{\partial u}{\partial \nu} - \frac{\partial v}{\partial \nu}=0, \quad x \in \partial \mathcal O,
\end{array}
\end{equation}
has a non-trivial solution. Under our assumptions, the set of ITEs is discrete \cite{sylv}.

 There is a wide literature on the inverse scattering where the scattering signal is registered in a specific direction. A classical example of this situation is the backscattering problem.
The uniqueness of solution of the inverse problem using the backscattering data can be found in \cite{er1},\cite{er2},\cite{er3},\cite{p}, and the recovery of singularities was studied in \cite{rakesh},\cite{small},\cite{paiv},\cite{ruiz}. In all the papers above, it was
assumed that the echo data is available for incident waves coming from all the directions.

%{\bf When scattering data is available at some surface soderzhashuyu $\mathcal O$ vnutri then inverse scattering problem at a fixed energy could be reduced to Gelfand-Calderon problem and is widely studied, see citation in \cite{b} and \cite{RN2}.}

There are important applications where an observer has access to the obstacle only from one side. Also the incident waves can be often emitted only from a bounded region, and not from
infinitely remote points as in the classical backscattering problem. Recently, such a potential scattering problem (i.e., problem (\ref{eq2}) with a potential that is smooth in $\mathbb R^3$) has been studied by Rakesh and Uhlmann \cite{rakesh2} in a non-stationary setting.
They assumed that the incident waves are emitted from points $x$ varying in some sphere. The authors show uniqueness for spherically symmetrical potentials. Christiansen  \cite{c} studied the backscattering obstacle problem when the scattering amplitude was known in a neighbourhood of the reflected ray. The author obtained a uniqueness result in the class of strictly convex obstacles.

In this article we consider the scattering problems (\ref{eq1}) and (\ref{eq2}) when the incident waves are emitted from a smooth surface $\obs$ that is a boundary of a bounded domain $B$ located outside of $\mathcal O$, i.e., $\overline{B}\bigcap\overline{\mathcal O}=\emptyset$. We assume that the receivers are also distributed over the same surface $\obs$, i.e., the following data is available:
\begin{equation}\label{6decA}
\left \{ \sol|_{\obs} : ~ u^{inc} \mbox{  emitted from  } \obs \right  \}
\end{equation}
The key element of our approach is Lemma \ref{6DecD} which states that restrictions of  waves emitted from $\mathcal S$ form a dense set in $L_2(\partial \mathcal O)$ due to the fact that $\obs$ is closed.

We will show that data (\ref{6decA}) for values of $\lambda$ in some interval $\mathcal I\in R^+$ allows one to determine effectively the interior eigenvalues of the scatterer that belong to the interval $\mathcal I$. As a byproduct, this leads to the uniqueness of the solution of the inverse problem when the data is given for all $\lambda>0$ and the spectrum defines the scatterer uniquely. The latter assumption is true for many classes of obstacles (see \cite{zeld}) and spherically symmetrical potentials (see \cite{mp}, \cite{chen}, \cite{cgh}). We believe that the arguments below allow for the reduction of the amount of scattering data in many methods used in inverse problems to identify properties of the scatterer, but here our aim is restricted to the recovery of the interior eigenvalues.

{\bf Acknowledgments.} The authors are thankful to Rakesh, D.Colton and A.Lechleiter  for useful discussions.

\section{The main result}
From now on, for the sake of simplicity of notations, we assume that $d=3$. The results and proofs in other dimensions $d\geq 2$ are similar. Define operators
$$
\mathcal L ~ : ~ L_2(\obs) \rightarrow H^1(\partial \mathcal O), \quad \mathcal L^* ~ : ~ L_2(\partial \mathcal O)
\rightarrow H^1(\obs) ,
$$
\begin{equation}\label{opL}
(\mathcal L \varphi)(x)= \int_{\obs} \frac{e^{-ik|x-y|}}{|x-y|} \varphi(y) dS_y, \quad
(\mathcal L^* \mu)(x)= \int_{\partial \mathcal O} \frac{e^{ik|x-y|}}{|x-y|} \mu(y) dS_y, \quad k=\sqrt\lambda>0,
\end{equation}
 where $H^s$ is the Sobolev space of functions on $\partial \mathcal O$ or $\obs$. We chose $s=1$, but could choose any $s\leq 2$. The upper bound is due to the restriction on the smoothness of $\partial \mathcal O$.

\begin{lemma}\label{6DecD}
Let $\lambda>0$ be not an eigenvalue of the negative Dirichlet Laplacian in either of the domains $\mathcal O$ or $B$ (with the boundary $\obs$).
Then operators $\mathcal L, \mathcal L^*$ have dense ranges. If $\partial \mathcal O$ or $\obs$ is infinitely smooth, then the range of $\mathcal L$ ($\mathcal L^*$ respectively) is dense in the Sobolev space $H^s(\partial\mathcal O)$ ($H^s(\mathcal S)$, respectively) with arbitrary $s\in \mathbb R$.
\end{lemma}

{\bf Proof.}
Let us prove that the range of $\mathcal L$ is dense.
Obviously, it is enough to show that the kernel of the operator $\mathcal L^* $ is trivial. Assume that the opposite is true. Then there exists $\mu=\mu(x),~x\in \partial \mathcal O$, such that $\mu\not\equiv 0$
and the function
$$
u:= \int_{\partial \mathcal O} \frac{e^{ik|x-y|}}{|x-y|} \mu(y) dS_{y}, \quad x\in \mathbb R^3,\quad k=\sqrt\lambda>0,
$$
which is defined on $\mathbb R^3$ and coincides with $\mathcal L^* \mu$ on $\obs$, vanishes on $\obs$. Since
\[
 (-\Delta-\lambda)u=0, \quad x\notin \partial \mathcal O,
\]
and $\lambda$ is not an eigenvalue of the Dirichlet problem in $B, ~u\equiv 0$ on $B$. Then from the equation above it follows that $u\equiv 0$ on $\mathbb R^3\setminus \mathcal O$. Thus $u$ satisfies the Helmholtz equation and the homogeneous Dirichlet boundary condition in $\mathcal O$. Since $\lambda $ is not an eigenvalue, it follows that $u\equiv 0$ in $\mathcal O$, i.e., $u\equiv 0$ in $\mathbb R^3$. The latter contradicts the fact that the jump of the normal derivative of $u$ on $\partial\mathcal O$ is equal to $-4\pi\mu\not\equiv 0$. Thus the density of the range of the operator $\mathcal L$ is proved. Similar arguments are valid for $\mathcal L^*$.

 \qed

Consider the near field operator
$$
F_S =F_S(\lambda) : ~ L_2(\obs)\rightarrow H^1(\obs), \quad  F_S\varphi = \sol|_{\obs}, ~~\varphi\in L_2(\obs), \quad  \lambda\notin D,
$$
where $\sol$ is the solution of (\ref{eq1}) or (\ref{eq2})  with $u^{inc}=\mathcal L \varphi$ on $\partial \mathcal O$ (with $u^{inc}=\mathcal L \varphi$ on $\mathcal O$ in the case of transmission problem (\ref{eq2})).

 The main result below (Theorem \ref{tt1}) uses the non-physical operator $F_S(\lambda)$ outside of a discrete set of values of $\lambda$. Discrete sets of $\lambda$ are non-essential since the inside/outside duality method determines if a point $\lambda_0$ is an ITE based on the behavior of $F_S(\lambda)$ in a neighborhood of $\lambda_0$, not at $\lambda_0$. The operator $F_S(\lambda)$ is non-physical since $\mathcal L\varphi$ represents waves coming to $\obs$ from infinity, while waves emitted from $\obs$ have the form $\overline{\mathcal L}\varphi$. However, one can use operator $F_S(\lambda)$ since each function $\sol|_{\obs}=F_S\varphi$ can be obtained (and measured) using approximations by scattered fields on $\obs $ produced by some waves emitted from $\obs$. Namely, the following lemma holds.

%Note that operator $\mathcal L$ represents waves coming to $\obs$ from infinity, while waves emitted from $\obs$ with a source density $\psi$ have the form $\overline{\mathcal L}\psi$. Thus $F_S\varphi$ is not the scattered wave produced by sources on $\obs $ with the density $\varphi$.  However, $\sol|_{\obs}=F_S\varphi$ can be obtained (and measured) as a scattered field on $\obs $ produced by some waves emitted from $\obs$. Namely, the following lemma holds.
\begin{lemma}\label{11J1}
Let $\lambda>0$ be not an eigenvalue of the negative Dirichlet Laplacian in either of the domains $\mathcal O$ or $B$. Then  for each $\varphi\in L_2(\obs)$, one can construct a sequence $\psi_n\in L_2(\obs)$ of the source densities such that $F_S\varphi=\lim_{n\to 0}\sol_n|_{\obs}$, where the limit is taken in the space $ L_2(\obs)$ and $\sol_n$ is the solution of (\ref{eq1}) or (\ref{eq2})  with $u^{inc}=\overline{\mathcal L}\psi_n$. One can determine the source densities $\psi_n$ without a priori knowledge of $\mathcal O$ if the radius of a ball $|x|<1/\varepsilon$ containing $\mathcal O$ is known.
\end{lemma}
{\bf Proof.} Consider a bounded domain $\widetilde{\mathcal O}$ that contains $\overline{\mathcal O}$  and  such that ${\rm dist}(B, \widetilde{\mathcal O})>0$. For example, one can take $\widetilde{\mathcal O}=(\mathbb R^d\backslash\overline{ B_\varepsilon})\bigcap\{|x|<1/\varepsilon\}$, where
$B_\varepsilon$ is the $\varepsilon$-extension of $B$ and $\varepsilon>0$ is small enough. Without loss of generality, we can assume that the boundary of $\widetilde{\mathcal O}$ is infinitely smooth and $\lambda$ is not an eigenvalue of the negative Dirichlet Laplacian in $\widetilde{\mathcal O}$.

From Lemma \ref{6DecD} it follows that the range of the operator
$$
(\widetilde{\mathcal L} \varphi)(x)= \int_{\obs} \frac{e^{-ik|x-y|}}{|x-y|} \varphi(y) dS_y, \quad x \in \partial \widetilde{\mathcal O}, \quad   \varphi\in L_2(\obs),
$$
is dense in $H^{3/2}(\partial \widetilde {\mathcal O})$. Then the same is true for $\overline{\widetilde{\mathcal L}}$. Hence for every $\varphi \in L_2(\obs)$, there exists a sequence $\psi_n \in L_2(\obs)$ such that $\overline{\widetilde{\mathcal L}}\psi_n\to \widetilde{\mathcal L}\varphi$ in $H^{3/2}(\partial \widetilde {\mathcal O})$. Below we consider functions $\overline{\widetilde{\mathcal L}}\psi_n, \widetilde{\mathcal L}\varphi, \overline{\mathcal L}\psi_n, \mathcal L\varphi$ defined by the corresponding integrals for all $x\in\mathbb R^3$. The standard a priory estimate (e.g., \cite{mcl}) for the solution $u=\overline{\widetilde{\mathcal L}}\psi_n- \widetilde{\mathcal L}\varphi$ of the Helmholtz equation in $\widetilde{O}$ implies that
$$
\|\overline{\widetilde{\mathcal L}}\psi_n- \widetilde{\mathcal L}\varphi\|_{H^{2}(\widetilde{\mathcal O})} \leq C(\lambda) \|\overline{\widetilde{\mathcal L}}\psi_n- \widetilde{\mathcal L}\varphi\|_{H^{3/2}(\partial \widetilde{\mathcal O})}\to 0  \quad  {\rm as} \quad n\to\infty.
$$
Since $\mathcal O \subset \widetilde{\mathcal O}$, we have that
$$
\|\mathcal L \varphi - \overline{\mathcal L} \psi_n \|_{H^{3/2}(\partial \mathcal O)}\to 0  \quad  {\rm as} \quad n\to\infty
$$
and
$$
\|\mathcal L \varphi - \overline{\mathcal L} \psi_n \|_{H^2( \mathcal O)}\to 0  \quad  {\rm as} \quad n\to\infty.
$$
The statement of the lemma is an immediate consequence of the last two relations and a priory estimates (e.g., \cite{mcl}) for the solutions of problems (\ref{eq1}) and (\ref{eq2}) (with radiation condition at infinity).

\qed

Let us define the constant $\sigma$ as follows: $\sigma=1$ in the case of the Dirichlet boundary condition in the obstacle scattering problem and when $n(x)<1, x \in \partial \mathcal O,$ in the transmission scattering problem. Let $\sigma = -1$ in the case of the Neumann boundary condition and when $n(x)>1, x \in \partial \mathcal O$.

For each function $\varphi \in L_2(\obs)$, consider the complex number $(\sigma F_S\varphi,\varphi)$ and its argument in the interval $[0,2\pi)$. Let us introduce the function
\begin{equation}\label{new1}
\Phi(\lambda) = \inf_{\varphi \in L_2(\obs), ~ \varphi \neq 0} \arg(\sigma F_S(\lambda) \varphi,\varphi) \in [0,2\pi).
\end{equation}
The following theorem holds:
\begin{theorem}\label{tt1}
Operator $F_S(\lambda)$ given for all $\lambda \in \mathcal I \subset \mathbb R^+$, except possibly a discrete set of points, determines all the interior eigenvalues in the interval $\mathcal I$. Namely, $\lambda_0\in \mathcal I$ is an interior eigenvalue of an obstacle scattering problem if and only if
\begin{equation}\label{lim}
\liminf_{\lambda \rightarrow \lambda_0} \Phi(\lambda) = 0,~~\sigma=1, \quad
\liminf_{\lambda \rightarrow \lambda_0}[2\pi- \Phi(\lambda)] = 0,~~\sigma=-1,
\end{equation}
 where $\lambda\neq \lambda_0$. The same criterion is valid for the interior transmission eigenvalues of odd multiplicity (in particular, for simple ones).
\end{theorem}

 {\bf Remarks.} 1) Our proof is based on the inside-outside duality principle \cite{ds}, \cite{Smil}, \cite{ep}, \cite{EP2}, \cite{lvrem}, \cite{kl}, \cite{lv2014} and ideas developed in the study of denseness of the far-field patterns \cite{k1986}. These principles were used recently \cite{lech1},\cite{lech2},\cite{lech3} for numeric evaluation of the interior eigenvalues  for obstacle scattering, inhomogeneous medium scattering and electromagnetic scattering.

 2) In the case of the obstacle scattering, one can easily refine the statement of the theorem and find the multiplicity of the eigenvalues. This can not be done for the transmission problem where the inside-outside duality principle allows one to observe not all the multiple eigenvalues (with a fixed $\lambda$), but only some of them, and an even number of them can be missed (see more details in \cite{lv2014}). One also can consider one sided limits in (\ref{lim}) in the case of the obstacle scattering: $\lambda \rightarrow \lambda_0$ can be replaced by
$\lambda \uparrow \lambda_0$ if $\sigma=1$, and by $\lambda \downarrow \lambda_0$ if $\sigma=-1$.

3) While the proof of the theorem uses the lemma below that contains an assumption on the location of the spectrum of the negative Dirichlet Laplacian, the theorem is proved without a priori assumptions on this spectrum since criterium (\ref{lim}) does not require the knowledge of the operator $F_S(\lambda)$ at all the points $\lambda$. It is assumed there that $\lambda\in\mathcal I$ with an arbitrary discrete set being omitted.

4) We believe that Theorem \ref{tt1} holds when the boundary $\partial \mathcal O$ is Lipshitz. We require $C^2$ boundary in order to avoid discussions on existence and properties of Dirichlet-to-Neumann maps.

The following lemmas will be needed to prove the theorem above.

Denote by
$$
F_1(\lambda),F^{out}(\lambda): H^1(\partial\mathcal O)\to L_2(\partial\mathcal O)
$$
the Dirichlet-to-Neumann maps for the Helmholtz equation in the interior and the exterior of $ \mathcal O$, respectively. The solutions are assumed to satisfy the radiation condition when $F^{out}$ is defined. It is well known that these operators are pseudo differential operators of order one (see \cite{vb1}-\cite{vb2}) and bounded from $H^s$ to $H^{s-1}, s\leq 2$. The restriction on $s$ is not needed if $\partial\mathcal O$  is infinitely smooth. When $\lambda$ is positive, operator $F^{out}$ is analytic in  $\lambda$, and  $F_{1}$ has poles at eigenvalues of the negative Laplacian in $\mathcal O$. The Dirichlet-to-Neumann map $F_n$ for the equation $(\Delta+\lambda n)u=0$ in $ \mathcal O$ is defined similarly.  The normal vector in all the cases is assumed to be directed outside of $\mathcal O$.
\begin{lemma}\label{lsc}
Operator $F_S(\lambda),~\lambda>0,$ has the following representation:
\begin{equation}\label{6DecC}
F_S= \frac{-1}{4\pi} \mathcal L^* (F_1-F^{out}) \mathcal L
\end{equation}
in the case of the Dirichlet obstacle scattering,
\begin{equation}\label{mid}
F_S= \frac{1}{4\pi} \mathcal  L^* F_1((F_1)^{-1}-(F^{out})^{-1} ) F_1 \mathcal L
\end{equation}
in the case of the Neumann obstacle scattering, and
\begin{equation}\label{6DecC1}
F_S=\frac{1}{4\pi}\mathcal L^* (  F_1-F^{out}) (F_n - F^{out})^{-1}(F_1 -F_n)\mathcal L
\end{equation}
in the case of the transmission scattering problem.
\end{lemma}
{\bf Remarks.} 1) These formulas are direct analogues of the formulas for the scattering amplitude in the problem of scattering of the plane waves (see \cite{ep} for the obstacle problem and \cite[Th.2.3]{lv2014}  for the transmission problem). The only difference is that a plane wave is defined by the direction $\omega$ of the incident wave, and $\obs$ is replaced by the unit sphere $S^2=\{\omega: |\omega|=1\}$ in that case. The operators $\mathcal L, \mathcal L^*$ are also slightly different in the case of the plane waves. We will denote them by $\widehat{\mathcal L}, ~\widehat{\mathcal L}^*$. Then
\begin{equation}\label{opL1}
\widehat{\mathcal L} :L_2(S^2) \rightarrow L_2(\partial \mathcal O), \quad \widehat{\mathcal L} \varphi(x)= \int_{S^2}e^{ik\omega\cdot x} \varphi(\omega) dS_\omega.
\end{equation}

2) Note that operator $(F_n - F^{out})^{-1}$ is analytic in $\lambda$ for $\lambda>0$, see \cite{lv2014}, Lemma 2.1 item 2. Thus, the right-hand sides in (\ref{6DecC})-(\ref{6DecC1}) are meromorphic in $\lambda$ when $\lambda>0$. Operator $F_S(\lambda)$ is analytic there since it is defined via the solution of the exterior problem with an analytic Dirichlet boundary condition, and the latter problem does not have eigenvalues embedded into continuous spectrum. Hence if relations (\ref{6DecC})-(\ref{6DecC1}) are established for all $\lambda>0$ except possibly a discrete set, then they can be extended by continuity toq all $\lambda>0$.

{\bf Proof}. Let us prove (\ref{6DecC}). Note that  $u^{inc}=\mathcal L \varphi$. We will look for $u^{sc}$ in the form of the potential $u^{sc}=\mathcal L^*\mu$ with an unknown density $\mu$. Since $u^{sc}=-\mathcal L \varphi$ on $\partial \mathcal O$, it follows that
\[
F^{out}u^{sc}=-F^{out}\mathcal L \varphi, \quad   F_1u^{sc}=-F_1\mathcal L \varphi.
\]
Since the jump on $\partial \mathcal O$ of the normal derivative of the potential $\mathcal L^*\mu$ is equal to $-4\pi\mu$, from the formulas above it follows that $\mu=\frac{-1}{4\pi}(F_1-F^{out})\mathcal L \varphi$. It remains only to put this expression into $F_S\varphi =u^{sc}=\mathcal L^*\mu, ~x\in \obs$.

The case of the Neumann condition is treated similarly using the Neumann-to-Dirichlet map instead of the Dirichlet-to-Neumann map. The proof of  (\ref{6DecC1}) is similar to the proof of the same formula  \cite[Th.2.3]{lv2014} in the case of plane incident waves. We will not repeat it here since it is somewhat cumbersome.
\qed

Let  $F(k):L_2(S^2)\to L_2(S^2)$ be the far-field operator, i.e., the integral operator whose kernel $f(\theta,\omega)$ is the scattering amplitude $u^{sc}_\infty(k,\theta)$ (see (\ref{rc})) defined by the plane incident wave  $u^{inc}=e^{ikx\cdot \omega}$.
\begin{lemma}\label{6DecB}
Let $\lambda>0$ be not an eigenvalue of the negative Dirichlet Laplacian in either of the domains $\mathcal O$ or $B$. Then the following sets coincide
$$
\overline{\{\arg(F(k)\varphi,\varphi),~ \varphi \in L_2(S^2)\}}= \overline{\{\arg(F_S(\lambda)\varphi,\varphi), ~ \varphi \in L_2(\obs)\}}, \quad \lambda=k^2,
$$
where $F(k)$ is the far-field operator.
\end{lemma}
{\bf Proof.} We will give a proof only for the Dirichlet obstacle scattering since the other cases are treated absolutely similarly. Due to Lemma \ref{lsc}, we have
$$
(F_S\varphi,\varphi)= ((F^{out}-F_1)\mu,\mu), \quad \mu=\mathcal L \varphi,
$$
where $\mathcal L$ is defined by (\ref{opL}). From the remark following Lemma \ref{lsc}, it follows that
$$
(F(k)\varphi,\varphi)= ((F^{out}-F_1)\mu,\mu), \quad \mu=\widehat{\mathcal L} \varphi,
$$
 where $\widehat{\mathcal L}$ is defined by (\ref{opL1}).
It remains to note that operators $\mathcal L$ and $\widehat{\mathcal L}$ have dense ranges.

\qed

{\bf Proof of the main theorem.} Recall that the scattering matrix $S(k)$ is a unitary operator in $L_2(S^2)$ of the form
\begin{equation}\label{end1}
S(k)=I+\frac{ik}{4\pi} F(k).
\end{equation}
Its eigenvalues $z=z_j(k)$ belong to the unit circle. Operator $F(k)$ is compact, and its eigenvalues converge to the origin. Thus the eigenvalues $z_j(k)$ of $S(k)$ converge to the point $z=1$ when $k$ is fixed and $j\to \infty$. Functions $z_j(k)$ are analytic in $k\in \mathbb R^+$ everywhere except the essential point of the spectrum $z=1$.

We will need the inside-outside duality principle \cite{ds}, \cite{Smil}, \cite{ep}, \cite{EP2}, \cite{kl}, \cite{lv2014}, which concerns the behavior of the set of points $\{z_j(k)\}$ when $k$ changes from zero to $\infty$ passing through different points $k_0$ for which $\lambda_0=k_0^2$
 are interior eigenvalues.
 %Denote by $\gamma_{\alpha_1,\alpha_2}$ the arc $\alpha_1<\arg\lambda<\alpha_2$ of the unit circle in the complex $\lambda$-plane.
The principle states (omitting some details that are non-essential for our purpose) that {\it 1) one half of the unit circle (the upper part if $\sigma=1$, and the lower part if $\sigma=-1$) contains at most a finite number of the eigenvalues $z_j(k)$ of $S(k)$ for each fixed value of $k$,  and 2) if $m\geq 0$ is the multiplicity of an interior eigenvalue $\lambda=\lambda_0=k_0^2$,  then $m_1$ eigenvalues $z_j(k)$ of $S(k)$ from the half circle indicated above approach the point $z=1$ when $k\uparrow k_0$, and $ m_2$ eigenvalues $z_j(k)$ from the same half circle approach the point $z=1$ when $k\downarrow k_0$, where $m_1+m_2\leq m$} (see \cite[Lemmas 3.1,3.2]{lv2014} for the duality principle in the case of the transmission problem with the assumptions on the refractive index imposed in the present paper).

%Let us state it more rigorously in the
%case of eigenvalues of odd multiplicity :

%{\it Let $\sigma=1$.

%If $\lambda_0$ is not an interior eigenvalue, then there is an arc $\gamma_{0,\delta}$ free of eigenvalues of the $S(k)$ when $|\lambda - \lambda_0|<<1$.

%If $\lambda_0$ is an interior eigenvalue of odd multiplicity, then for any $\delta >0$, there exist $\delta_1>0$ such that $\gamma_{0,\delta}$ contains some eigenvalues of the $S(k)$ when either $\lambda \in (\lambda_0,\lambda_0+\delta_1)$ or $\lambda \in (\lambda_0 - \delta_1,\lambda_0)$.

%{\bf The same statements hold if $\sigma=-1$ and  $\gamma_{0,\delta}$ is replaced by  $\gamma_{-\delta,0}.$
%}}

These statements can be made more precise in the case of obstacle scattering. Namely,  $m_1=m,~m_2=0$ in the case of the Dirichlet boundary condition, and $m_1=0,~m_2=m$ in the case of the Neumann boundary condition. In general, these refined statements are not true for the ITEs, see \cite{lv2014}, but $m_1+m_2=m$ by modulus $2$ in the latter case.

Since operators $S(k)$ and $F(k)$ have the same eigenfunctions with the eigenvalues $z_j(k),~\chi_j(k) $, respectively, and
\begin{equation}\label{end}
z_j(k)=1+\frac{ik}{4\pi}\chi_j(k) ,
\end{equation}
the inside-outside duality principle can be reformulated in terms of the far field operator  $F(k)$  instead of the scattering matrix $S(k)$.
Namely, from $|z_j(k)|=1$ and (\ref{end}) it follows that one can replace condition $z_j\to1$ by either $\arg \chi_j(k)\downarrow 0$ if $\sigma=1$ (and $z_j(k)$ belong to the upper half circle) or by $\arg \chi_j(k)\uparrow \pi$ if $\sigma=-1$ (and $z_j(k)$ belong to the lower half circle).

Next, consider the quadratic form $(\sigma F\varphi,\varphi)$ and choose  $\arg(\sigma F\varphi,\varphi)\in[0,2\pi)$. Then the convergence of $\arg \chi_j(k)$ can be replaced by the requirement of the existence of finite-dimensional subspaces on which $\arg(\sigma F\varphi,\varphi)$ converges to $0$ when $\sigma=1$ or to $2\pi$ when $\sigma=-1$. For the sake of simplicity, we assume below that $\sigma=1$. The case of $\sigma=-1$ is treated similarly. Then the inside-outside duality principle implies that $\lambda=\lambda_0=k_0^2$ is an interior eigenvalue of multiplicity $m\geq 0$ if and only if there exist $m_1~(m_2)$-dimensional subspaces $\Phi_{1}~ (\Phi_{2})$ on which $\arg(F\varphi,\varphi)\downarrow 0$ when $k\uparrow k_0$ ($k\downarrow k_0),$ respectively. The transition from $\chi_j$ to the quadratic form is obvious: one can choose $\Phi_i$ to be the span of the corresponding eigenfunctions of operator $F$. The inverse transition is based on simple geometrical arguments which follow from (\ref{end1}) and unitarity of $S(k)$. These arguments can be found in \cite[Lemmas 3.1,3.2]{lv2014}.  The paper \cite{lv2014} is devoted to the transmission problem, but these lemmas have a general character and are applicable to the scattering by an obstacle as well. It is worth mentioning that the proof of the inside-outside duality principle is often based on establishing the corresponding relation for $\arg(\sigma F\varphi,\varphi)$, and then the statement of the principle in terms of eigenvalues of $S(k)$ follows from here \cite{lv2014},\cite{kl},\cite{lech3}.

Due to Lemma \ref{6DecB}, one can replace $F$ in the quadratic form $\arg(\sigma F\varphi,\varphi)$ by $F_S$. Thus Theorem \ref{tt1} is an immediate consequence of the inside-outside duality principle (stated in terms of the quadratic form of $F$) and Lemma \ref{6DecB}.

\qed

\end{document}